\documentclass[fleqn,10pt]{wlscirep}
\usepackage[utf8]{inputenc}
\usepackage[T1]{fontenc}

\usepackage{lineno}
\usepackage{booktabs}
\usepackage{multirow}
\usepackage{longtable}
\usepackage{xr-hyper}
\usepackage[charter,cal]{mathdesign}

\makeatletter
\newcommand*{\addFileDependency}[1]{
  \typeout{(#1)}
  \@addtofilelist{#1}
  \IfFileExists{#1}{}{\typeout{No file #1.}}
}
\makeatother

\newcommand*{\myexternaldocument}[1]{%
    \externaldocument{#1}%
    \addFileDependency{#1.tex}%
    \addFileDependency{#1.aux}%
}
\myexternaldocument{si}

\title{
Remote sensing and GPS mobility reveal heat's impact on human activity across diverse climates
}

\author[a,*]{Andrew Renninger}
\author[b]{Olena Holubowska}
\author[c]{Paul Blanchard}

\affil[a]{Centre for Advanced Spatial Analysis, University College London, London, UK}
\affil[b]{Department of Earth and Environmental Sciences, KU Leuven University, Leuven, Belgium}
\affil[c]{Development Impact Group, World Bank, Washington, DC, US}

\affil[*]{Corresponding author: Andrew Renninger (E-mail: andrew.renninger.12@ucl.ac.uk)}

\begin{abstract}
Extreme heat is a growing threat to both individual livelihoods and broader economies, killing a growing number of people each year as temperatures rise in many parts of the world and limiting productivity. Many studies document the link between heat waves and mortality or morbidity, and others explore the economic consequences of them, but few are able to determine how populations respond to the shock of extreme heat in day-to-day activity. Toward this end, we investigate the link between human mobility and ambient temperature. Examining Indonesia, India and Mexico, we show that extreme heat reduces mobility by up to 10\% in urban settings, with losses concentrated midday. We examine the shape of the relationship, finding that while heat reduces activity, very hot days and very long heat waves may induce more of it, indicating different adaptation. Effects are stronger in poorer areas. Twinning these models with climate projections, we show that without adaptation mobility may fall 1-2\% per year on aggregate, with certain seasons and places seeing activity fall by as much as 10\%. According to our estimates, small cities will face the highest relative losses and large cities will experience the greatest absolute impacts.
\end{abstract}
\begin{document}

\flushbottom
\maketitle

\section*{Introduction}
Extreme heat threatens lives globally \cite{zhao2024global, masselot2023excess}, with rising temperatures linked to greater mortality \cite{kaiser2007effect} and morbidity \cite{zhang2015impact}. Vulnerable groups, including the very young and old, suffer the worst outcomes \cite{zhao2024global, kenny2010heat}. Beyond human health, heat waves induce economic damage by lowering productivity \cite{somanathan2021impact} and slowing growth \cite{dell2014we}. The severity of this challenge continues to mount: heat waves are intensifying in both magnitude and frequency \cite{perkins2012increasing}, with their duration growing since 1950 and that rate of growth accelerating since 1980 \cite{perkins2020increasing}. Yet, the behavioral dimensions of heat impacts are not clear and research is often limited by context, often focusing on advanced economies \cite{pappalardo2023future}. By pairing weather and socio-economic data with GPS mobility, we show that extreme heat fundamentally disrupts activity urban areas. In doing so, we demonstrate a scalable approach for modeling human adaptation to extreme heat across diverse contexts. 

The consequences of extreme heat differ based on who people are and where they live. Age, gender, and surrounding conditions shape vulnerability \cite{xu2014impact, oudin2011heat}, while geographical features like water bodies \cite{burkart2016modification}, urban design \cite{huang2024emergence}, and broader geographic characteristics \cite{zschenderlein2019processes} alter heat resilience. Socioeconomic status also matters: in areas with abundant air conditioning or car ownership, heat waves pose fewer disruptions \cite{gu2024socio}, whereas disadvantaged communities experience sharper mobility shifts due to hot weather \cite{gu2024socio}. In the US, wealth reduces heat-related mortality via air conditioning \cite{barreca2016adapting}, as people pay more to avoid heat than cold \cite{albouy2016climate}, cut working hours, and redirect outdoor leisure indoors \cite{graff2014temperature}. Moreover, although wealthier populations can better delay activities, poorer ones resume hazardous routines sooner, facing heightened exposure \cite{li2022spatiotemporal}.

Populations do adapt over time: mortality decreases with more frequent high temperatures, implying compensatory mechanisms \cite{wu2024temperature}, and the least deadly temperatures often match local climate norms \cite{yin2019mapping}. Governments now classify extreme heat alongside hurricanes and cyclones, going so far as to name heat waves \cite{metzger2024beliefs} and devise “heat action plans” \cite{hess2023public}. Yet adverse outcomes persist: extreme heat lowers productivity in manufacturing \cite{somanathan2021impact} and agriculture \cite{schlenker2009nonlinear}, dampens growth in poorer regions \cite{dell2014we}, and a single anomalously hot day can curb output \cite{zhang2018temperature}. Heat also contributes to conflict \cite{burke2015climate}, erodes learning in non–air-conditioned classrooms \cite{park2020heat}, and strains infrastructural \cite{ferranti2016heat} as well as natural systems \cite{teskey2015responses}. Consequently, understanding how people adjust their daily activities is crucial.

Heat exerts its effects via physiological stress—causing dehydration, fatigue, and organ damage, especially among older individuals \cite{mcmichael2011climate, hess2023public}—and by compromising cognitive performance \cite{danner2021simulated}. This leads to fewer working hours \cite{zander2015heat}, constrained productivity in physically demanding sectors such as construction \cite{romanello20212021}, and altered urban behavior. Rising temperatures reshape activity patterns in cities worldwide \cite{li2024urban, ly2023exploring}, driving individuals to forgo short trips \cite{gu2024socio}, shift destination choices \cite{horanont2013weather}, and reduce both public transport use \cite{ngo2019urban} and walking \cite{kumakura2024assessing}. These behavioral modifications, rooted in discomfort or safety concerns, underlie broader economic and social impacts. As climate change intensifies, deeper knowledge of how heat influences collective movement has become a research priority \cite{vargas2023prioritize}.

In this paper, we investigate how extreme heat modifies urban mobility in three developing countries—India, Indonesia, and Mexico—home to 1.8 billion people and spanning eight climate zones \cite{beck2018present}. We combine mass GPS-based mobility data with thermal comfort measures to quantify the form and magnitude of heat’s influence in 2019. By modeling changes in trips, distances, and destinations, we illuminate the behavioral foundations of adverse health and economic outcomes. Given that hotter regions suffer the greatest productivity losses under climate change \cite{cruz2021global}, understanding these responses is essential. We find that extreme temperatures significantly alter mobility patterns, with activity levels dropping by up to 10\% during heat waves. We further disaggregate our analysis by population density and socio-economic status, yielding comparative insights into how different groups manage and recover from extreme heat across diverse climates and cultures, showing that economically disadvantaged areas show stronger reductions in mobility. While populations attempt to compensate by shifting activities to cooler hours, this adaptation is less effective in poorer areas, likely due to limited access to cooling infrastructure and inflexible work schedules. Our findings reveal how climate change could exacerbate existing inequalities in developing countries, where the most vulnerable populations face the greatest disruption to their daily activities. Since urban economies depend on the mobility and interaction \cite{duranton2004micro}, our results suggest that without interventions to help populations adapt to extreme heat, rising temperatures will impose a cost that falls most on those who can least afford it. 

\section*{Results}
\subsection*{Impact of hot days on mobility}
We use daily flows between origin and destination geohash5 aggregations, comprising $\sim236$ million trips and $\sim1.4$ billion unique GPS signals, gathered from mobile devices in the course of daily usage (for descriptive statistics and illustrations of the network, see Fig. \textbf{S1} and \textbf{S2} in the supplemental information). As these data best represent urban populations, we remove rural areas—clipping each network to ``functional urban areas'' \cite{schiavina2023fua}; Because the network is sparse and our concern is ``activity'', we define it as the number of journeys ending in each geohash5. For each geohash5, we acquire Universal Thermal Climate Index (UTCI) \cite{di2021era5} from ERA5-HEAT \cite{ERA5}, which takes wind, sun, moisture, and temperature into account to produce a measure of how hot or cold it feels. Joining the datasets, we are able to make inferences about the effect of temperature on moves terminating in a given geohash5. (See Methods for greater detail.)

\begin{figure*}[bt!]
\centering
\includegraphics[width=1\textwidth]{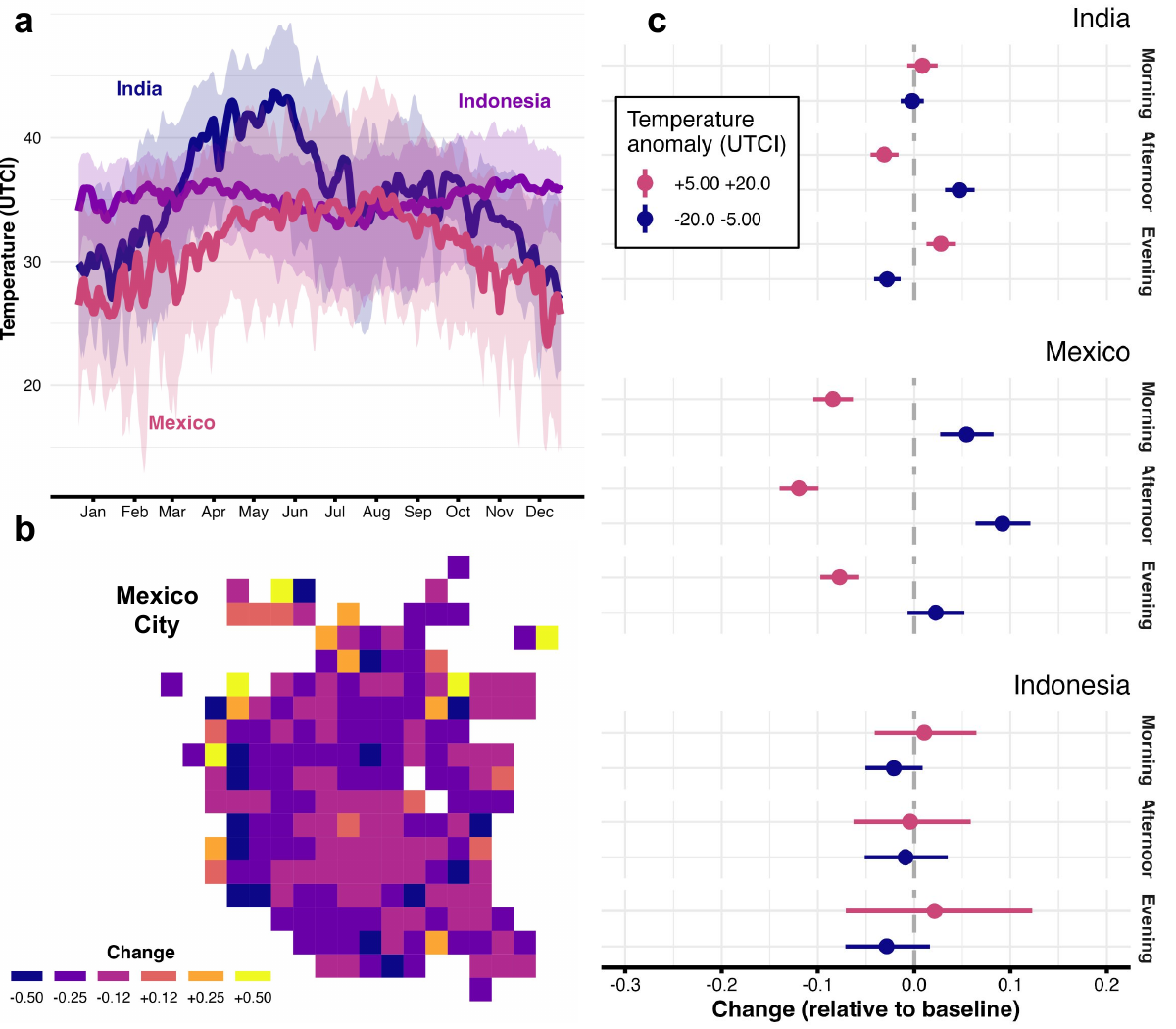}
\caption{\textbf{Extreme heat and mobility.} \textbf{A} Temperatures in India, Mexico and Indonesia throughout 2019, showing that we have three different climate profiles: equatorial—and thus stable—in Indonesia, and tropical but varied and seasonal in India and Mexico. \textbf{B} Change in mean daily visitation in Mexico City between a heat wave in July and the weeks preceding and following it; activity levels fell 10\% on aggregate, and although some peripheral areas gained foot traffic, losses are concentrated in the central city. \textbf{C} Results from a TWFE regression showing that, with controls for location and date, activity falls most in the afternoon on hot days—when the temperature is hottest—in Mexico and India. In Mexico activity falls throughout the day, while in India activity rises in the evening, after it has cooled off, on hotter days. Activity rises in the afternoon on cold days, suggesting that on cold days, the warmest hours of the day are conducive to activity, while on hot days, the hottest hours are destructive.}
\label{fig1}
\end{figure*}

We show temperature fluctuations across India, Mexico and Indonesia throughout the year in \textbf{\ref{fig1}A}, finding that each country gives us insights into different climates: located on the equator, Indonesia has a stable climate; Mexico and India have more pronounced curves with distinct profiles. Mexico City suffered heat wave in July 2019 (3+ days in the highest temperature decile \cite{perkins2013measurement}), and we show in \textbf{\ref{fig1}B} mean activity per grid cell during the heat wave compared to the weeks before and average, for comparison. Although noisy, activity falls in the center, with mixed effects in the periphery. With preliminary evidence that heat changes behavior, we use a TWFE approach to understand the effects across all countries systematically. Because frequent temperatures are less lethal \cite{wu2024temperature, yin2019mapping}, we examine deviation from the mean temperature in a geohash5. The results are shown in \textbf{\ref{fig1}C}, which plots the change relative to the strength of temperature anomaly. The results are clear in Mexico: higher temperatures mean fewer trips in a cell. In India, heat reduces afternoon activity—when temperatures peak—but activity shifts to the evening. Sensibly, the effect is inverted for cold, where waiting for the air to warm is better than during extreme heat. In Indonesia, there is limited variation and the effects we can observe are \emph{de minimis}, save for the possibility—given large standard errors—that cold appears to decrease activity throughout the day while heat appears to increase it.   

\subsection*{Insights into the shape of the temperature-mobility relationship}
To capture incremental temperature changes, we model the temperature–activity relationship using a generalized additive model (GAM), which offers non-parametric flexibility and parametric interpretability. Our GAM uses smooth functions that accommodate varying coefficients and suits the documented nonlinear temperature effects on productivity \cite{burke2015global}. (See Methods.) We use a directed acyclic graph (DAG) to identify the minimally sufficient adjustment set—the controls that allow us to infer causal links between our independent variable of interest (temperature) with our dependent variable (activity). For controls, we identify temporal and spatial factors as determinants of activity and use geohash5 fixed effects along with splines to capture trends over time. In alternative specifications, we interact temperature with population and deprivation to explore how these moderate its effects. While the GAM captures the shape of relationship, we use another TWFE specification with various temperature bins to provide more rigorous identification by accounting for both location-specific characteristics and time-varying factors.

\begin{figure*}[bt!]
\centering
\includegraphics[width=1\textwidth]{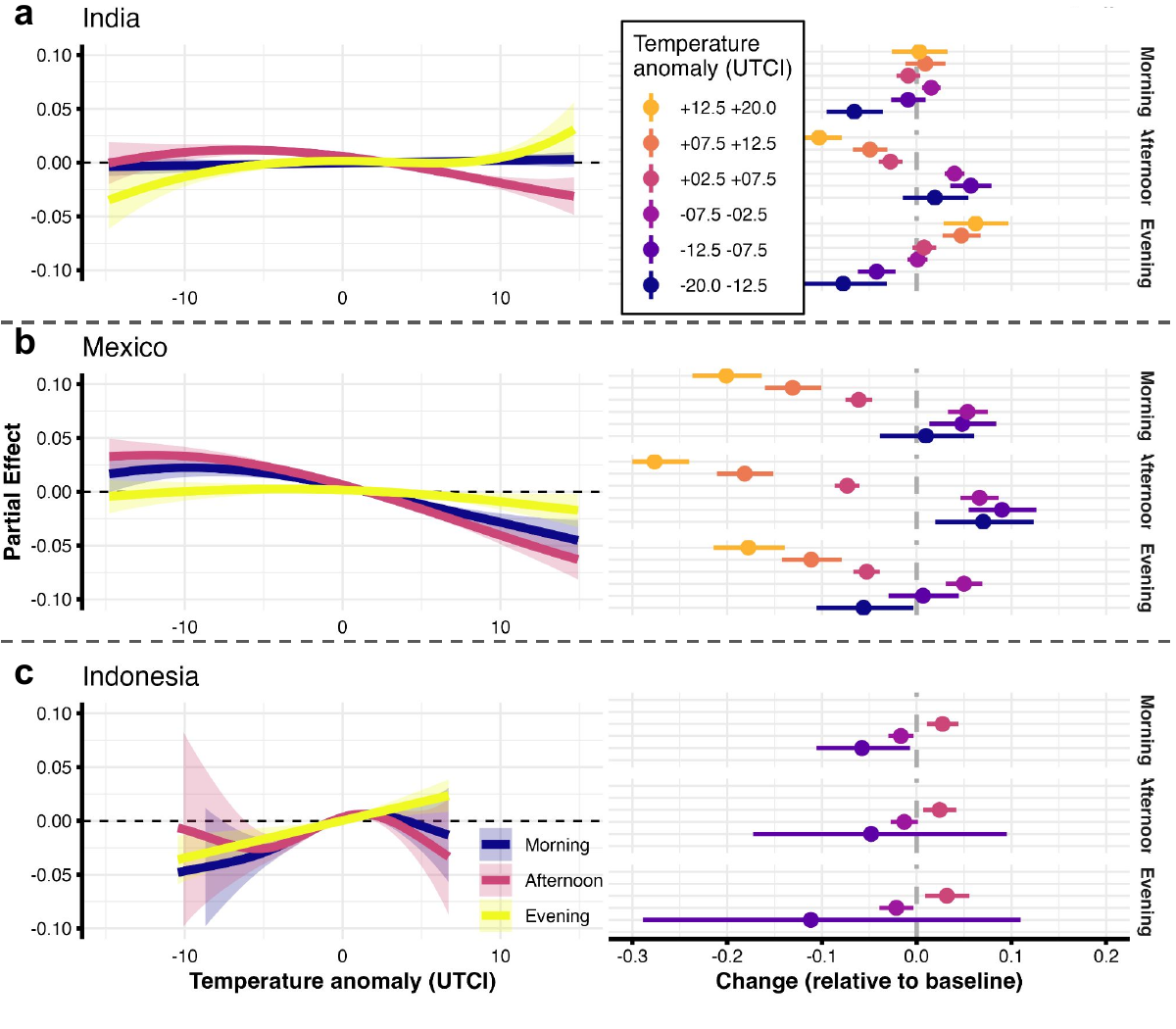}
\caption{\textbf{Modeling temperature and activity.} \textbf{A} We set up a directed acyclic graph (DAG) to ensure that we close all necessary causal paths to activity; we control for day-of-year, day-of-week, holidays, solar radiation, precipitation, and use geographic fixed effects that necessarily stratify by population and deprivation. \textbf{B} Model results for different countries, showing that the highest activity levels occur at average temperatures for an area, and that high extremes correspond with fewer trips. \textbf{C} Model results using trip duration rather than trip count reveal that the longest trips tend to occur at average temperatures, with extreme high temperatures leading to shorter trips.}
\label{fig2}
\end{figure*}

We compare both models in Fig. \textbf{\ref{fig2}}, finding consistent results. In Fig. \textbf{\ref{fig2}A} we see that in India, activity declines during the afternoon but recovers in the evenings. Notably, afternoon temperature peaks have inverse effects: most active on cold days, when warmth is beneficial, but least active on hot days. In Mexico, shown in \textbf{\ref{fig2}B}, higher temperatures are associated with fewer trips during all periods of the day. We see limited effects in Indonesia in Fig. \textbf{\ref{fig2}C}, with warmer temperatures increasing activity and cooler temperatures decreasing activity—though we cannot reject the null hypothesis here.    

Generally, deviation from a location’s mean temperature—hot or cold—reduces activity. Showing that relative changes in temperature are better predictors than absolute changes, our results are consistent with the literature that suggests climate preferences are set locally \cite{albouy2016climate}. This also agrees with research that finds that the most frequent temperature in an area is often the least fatal for that area's population \cite{yin2019mapping}. 

We also explore daily effects in the supplemental information. Again, we expect that it is relative changes rather than absolute values that matter, so we consider deviations from the average. Fig. \textbf{S3B} and \textbf{C} show the relationship between changes in temperature and trip counts and durations (measured in minutes) for full days. Our results agree with existing literature: as temperature increases relative to the local average, the number of trips falls. We also conduct placebo tests in \textbf{S4} and sensitivity analysis in \textbf{S5} to show that our results are unlikely to be spurious.

We then look at absolute temperatures to understand if levels are also a factor in driving activity. To see if prolonged heat waves differ, we test rolling averages of different lengths, where higher averages indicate stronger or longer fluctuations. Fig. \textbf{\ref{fig3}A} shows that in India, when the temperature exceeds $35^{\circ}$UTCI, the effect of temperature on activity becomes negative; Fig. \textbf{\ref{fig3}B} shows that in Mexico, this change from positive to negative occurs at $30^{\circ}$UTCI. In India, persistent heat waves (rolling averages above 
$45^{\circ}$UTCI) show relative adaptation, with effect sizes rising above zero. While not conclusive, this suggests that extended periods of high temperatures necessitate a return-to-activity, or that policy and behavior adapt only after multiple days of extreme heat. Replicating what we saw in the TWFE analysis, when we partition the data by time of day in Fig. \textbf{S6}, we see that the effect converging toward 0 at $42^{\circ}$UTCI in India at high temperatures is the product of countervailing forces: extreme heat increases activity in the evenings while decreasing it in afternoons, when the temperature is highest. This difference is only present in India, where temperature is more variable, but it indicates that compensatory changes are at play at the extremes. In Mexico, all times of day see reductions at high temperatures. 

\subsection*{Heterogeneities in the impact of heat on mobility}
\begin{figure*}[bt!]
\centering
\includegraphics[width=1\textwidth]{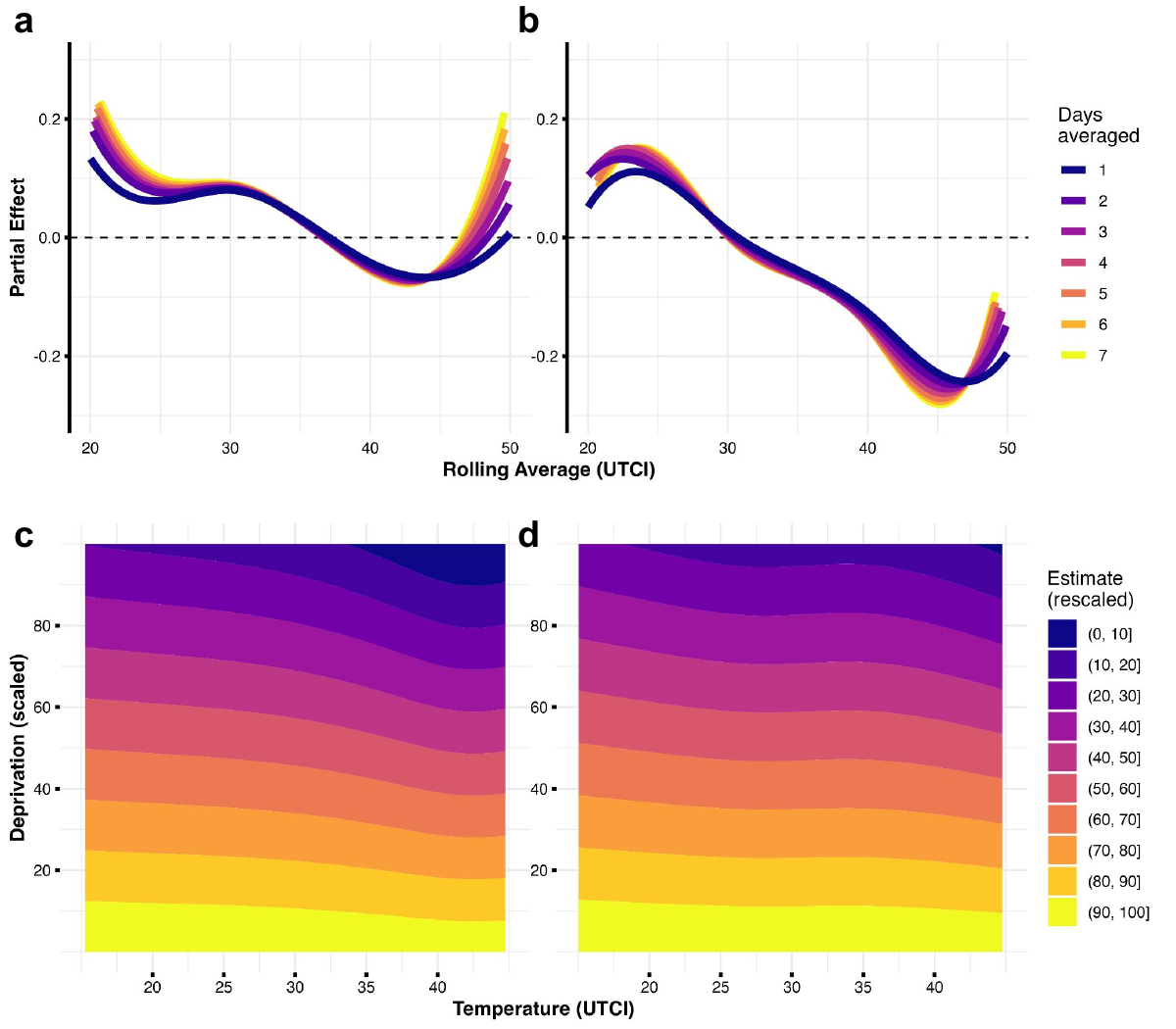}
\caption{\textbf{The effect of extreme heat on activity.} \textbf{A} Focusing on India, given that country's propensity for heat waves, we measure the duration and intensity of extreme heat by calculating a rolling average, finding that higher temperatures for longer are associated with stronger effects between $40-45^{\circ}$UTCI but not above that, which may indicate compensatory behavior at a certain point. \textbf{B} We fit the same model for Mexico and find similar results, with an ideal temperature and declining activity when heat surpasses it (noting that Mexico does not have the same extremes). Lasting heat waves, measured in rolling averages, show a stronger impact in Mexico and the rebound we see in India never becomes positive. \textbf{C} and \textbf{D} Looking at the interaction between deprivation and temperature in India and Mexico, respectively, we see a curve that follows above modeling for all socio-economic strata but it is steepest, indicating the strongest effects, for the poorest strata.}
\label{fig3}
\end{figure*}

We next examine how demographic and geographic factors shape activity. In this specification, we use geohash5 fixed effects but add interaction terms for deprivation \cite{ciesin} or population \cite{schiavina2023ghs}. In order to understand the interaction between temperature and deprivation, in particular, we explore the space of model predictions across all combinations of temperature and deprivation. These predictions are shown in Fig. \textbf{\ref{fig3}C} and \textbf{D}, for India and Mexico, respectively. (We use the same strategy for population in \text{S6}.)

In India, we recover the same arc as above, where activity partially recovers at the highest extremes. The contours in \textbf{\ref{fig3}C} are flatter for low deprivation and steeper for high deprivation, indicating that much of the worst effects of heat in India are felt by the poorest. Looking at Mexico in \textbf{\ref{fig3}D}, higher temperatures bring lower activity for all levels of deprivation, but the contours at higher levels of deprivation indicate that the effect is stronger for poor areas than for wealthy ones. We see recovery at extreme heat in India but not in Mexico, where contours steepen at higher temperatures. This has important implications for vulnerable populations, as we might expect competing forces to be at play: the poor are also less able to afford taking time off from work, but they may be less able to access cooling, which could suppress activity. We see evidence that the latter pressure dominates the former, and activity is suppressed at extremes.  

\begin{figure*}[bt!]
\centering
\includegraphics[width=1\textwidth]{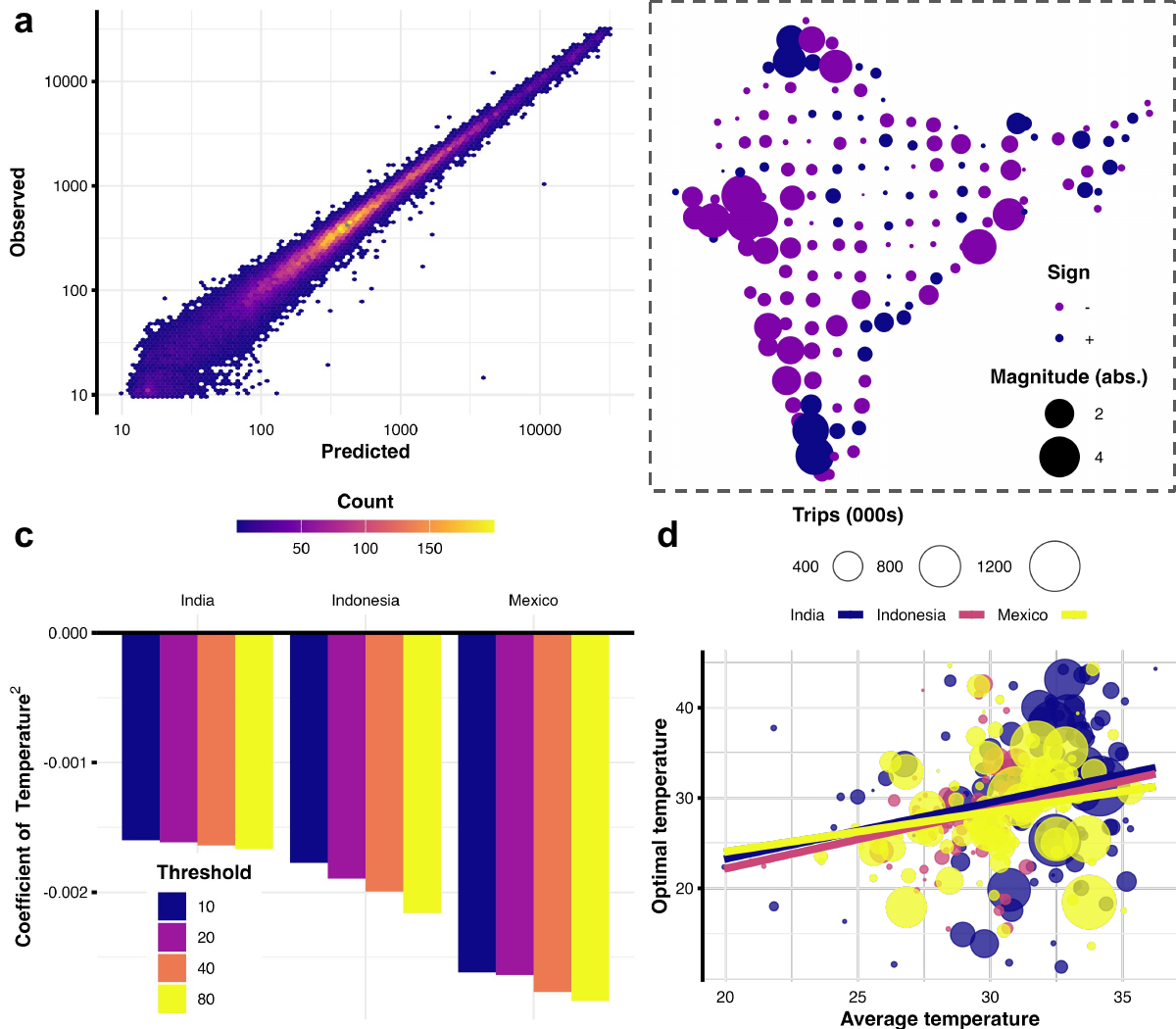}
\caption{\textbf{Alternative estimations across all countries.} \textbf{A} We fit an ARIMAX model to the time series of each geohash3, achieving good fit with a mean absolute percent error of 20\%. This model uses both autoregression—recent experience—and exogenous variation, temperature in our case, to predict activity, allowing us to interpret the effect of temperature; \textbf{B} shows the effect of $\text{Temperature}^2$ across India, with most areas showing a negative effect. \textbf{C} As a sensitivity analysis, we vary the required length of the time series, because some areas do not have complete data, and find that the effects only become stronger as we increase the minimum number of days in the series. Although a weak correlation of $\rho=0.25$, in \textbf{D} we see a relationship between average temperature and the optimal temperature for activity according to our model, supporting our earlier results.}
\label{fig4}
\end{figure*}

As a robustness check and to explore spatial variation, we use an ARIMAX model at the geohash3 level (to ensure full time series data for each). This model predicts activity by considering activity at time $t$ using activity at $t-1$ and $t-7$ along with linear and nonlinear terms for temperature, which allows the relationship to curve as temperature increases. This method implicitly controls for seasonality and day-of-week effects by using the previous day and the same day of the previous week to make a prediction, and achieves a good fit, shown for India in Fig. \textbf{\ref{fig4}A}. Important to this study, the quadratic term is negative across most regions, indicating that at a certain point the effect of temperature on activity turns negative. Fig. \textbf{\ref{fig4}B} shows, again for India, the sign and magnitude for each geohash3, which shows mixed results but less of temperature effect in the South, where temperatures are more consistent, and more of an effect in the North and Northwest. Some coefficients are positive with high magnitude, but these tend to be in the mountainous North, which does not have as many hot days, and the tropical South, which has less variability in temperature. We take the average coefficient for $\text{Temperature}^2$ across all countries in Fig. \textbf{\ref{fig4}C}. Note that we need to rescale the trip distribution in each country for the effect sizes to be comparable, so the value of each coefficient is difficult to interpret (in Fig. \textbf{S7}, we plot the curves from each ARIMAX formula to show how $\text{Temperature}^2$ influences the prediction). We filter out geohashes according to the length of the time series, because very short series could bias the results; generally, as this threshold of minimum observations rises, so too does the effect size.

Finally, we identify the optimal temperature by making predictions across the spectrum for each geohash3 and observing where the predictions begin to fall. According to these predictions, the mean and median optimal temperatures for India are $31.1^{\circ}$UTCI and $31.8^{\circ}$UTCI—in line with our estimates from the GAM. (These mirror what we see in \textbf{S7}, which plots the prediction curves for each geohash3 ARIMAX.) Although not a perfect match with the above findings, these results strengthen our assertion that extreme heat can limit activity. We document a noisy relationship between temperature optima across geohashes and the mean temperature of those geohashes ($\rho=0.25$ across all countries, $p<0.05$) in Fig. \textbf{\ref{fig4}D}: as mean temperature increases, the optimal temperature for activity also increases. This provides further evidence that populations adapt to climatic conditions. 

\subsection*{Projecting into the future}
We next turn to models of Earth's climate \cite{thrasher2022nasa} to make estimates of how activity will be impacted in future. In particular we use data from CMIP6 simulations, extracting the estimates from the Max Planck Institute because these estimates show the best fit to past trends \cite{craigmile2023comparing}. We substitute each 2019 temperature-day value with its 2050 counterpart from these climate models. We make these projections knowing that our own study suggests that optimal temperatures are often related to the climate of the area, and thus that the population of a given area may be able to adapt to local temperatures. Nevertheless rapid deviation from these temperatures as global warming accelerates \cite{perkins2020increasing} might still be a shock to communities who have acclimated to past temperatures. Furthermore, our data suggest that poor areas suffer the worst effects, and these areas might be the least able to cope with changes to the local climate going forward. As a modeling exercise, this provides us with a worst case, but a plausible one at that. 

\begin{figure*}[bt!]
\centering
\includegraphics[width=1\textwidth]{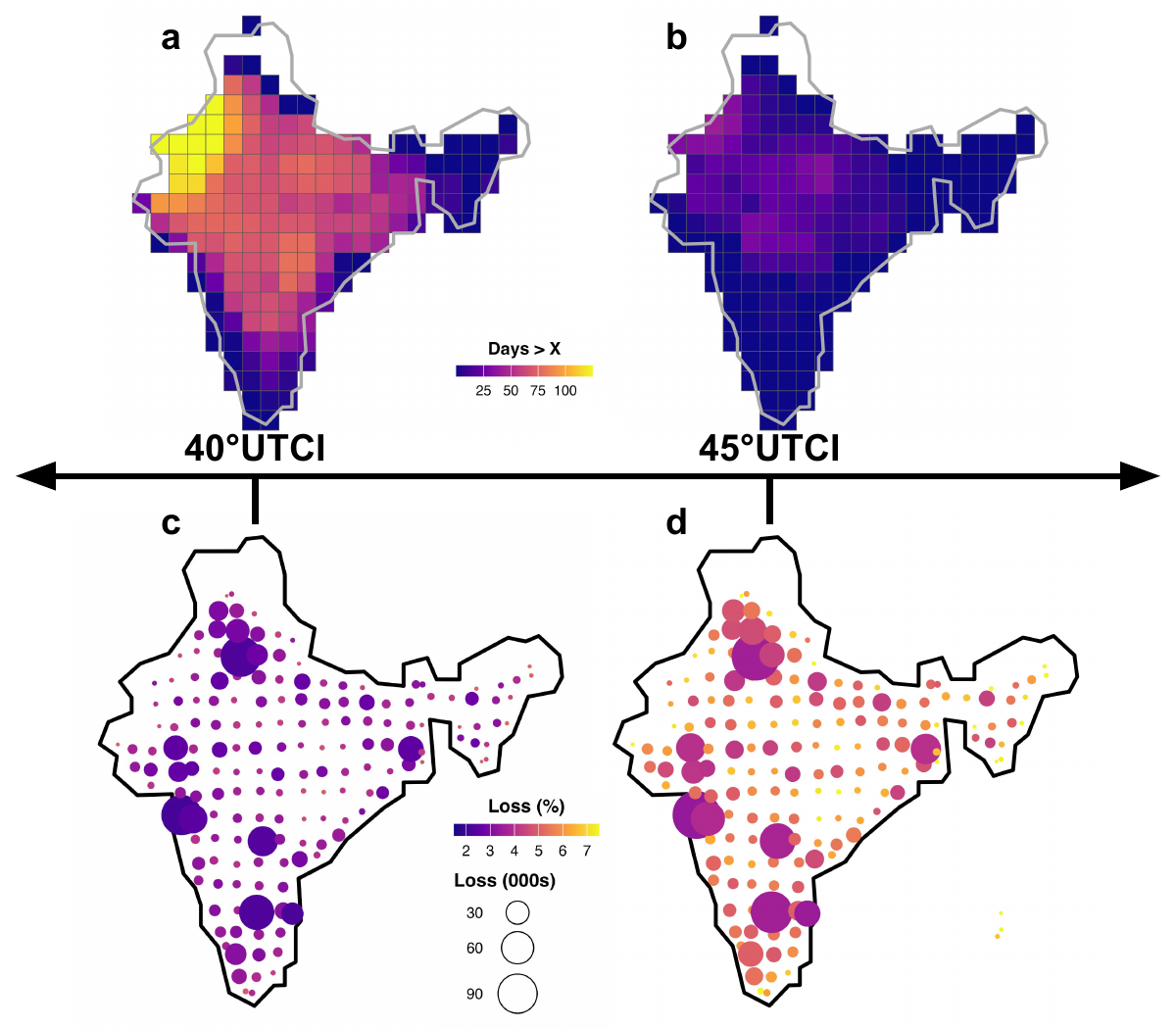}
\caption{\textbf{Future implications for India.} \textbf{A} CMIP6 projections for days above $40^{\circ}$UTCI in India by 2050; these indicate that Northern India could experience as many as 150 days per year at or above this temperature. \textbf{B} CMIP6 projects fewer days above $45^{\circ}$UTCI but much of the North will still experience 25 days at this extreme. \textbf{C} and \textbf{D} show the consequences of this heat according to our modeling: during those days between $40-45^{\circ}$UTCI, we can expect activity to decrease by $5\%$ in small cities and $2\%$ in large agglomerations; when the temperature is higher, the loss of activity could be as much as $7\%$ and $5\%$ in small and large urban areas respectively.}
\label{fig5}
\end{figure*}

CMIP6 projections (\textbf{\ref{fig5}A}) indicate 100–150 annual days above $40^{\circ}$UTCI in Northern states (e.g., Rajasthan), and (\textbf{\ref{fig5}B}) suggests 25 days above $45^{\circ}$UTCI. We distinguish between large urban areas (top 10 FUAs by population) and small cities (remaining FUAs). On any given day in June when such temperatures are most likely, not a weekend nor a holiday, our model estimates that a single day at $40^{\circ}$UTCI temperatures to reduce activity in large cities areas by 2\% and in small ones by 6\%, shown in Fig. \textbf{\ref{fig5}C}. Because megacities generate more activity, they experience a small relative change but a large absolute loss. As we see in Fig. \textbf{\ref{fig5}D}, these losses grow to 4\% and 8\% for large and small urban areas, respectively, during a single day of temperatures at $45^{\circ}$UTCI.

\begin{figure*}[bt!]
\centering
\includegraphics[width=1\textwidth]{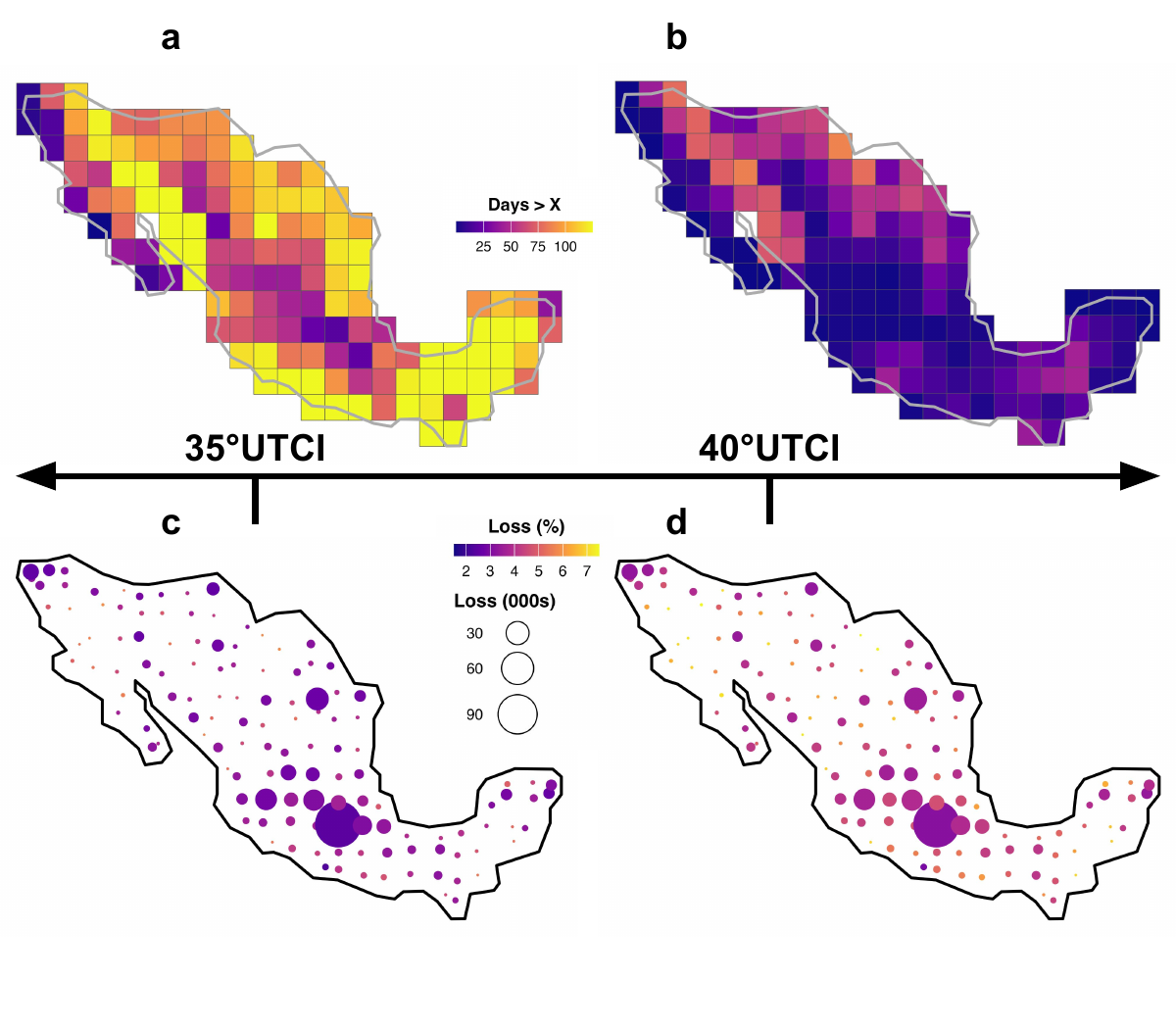}
\caption{\textbf{Future implications for Mexico.} \textbf{A} CMIP6 projections for days above a given threshold in Mexico by 2050, \textbf{A} and \textbf{B} show frequency of days above $35^{\circ}$UTCI and $40^{\circ}$UTCI, respectively; we see that the tropical South will have more frequent days above $35^{\circ}$UTCI but the arid North will have more days above $40^{\circ}$UTCI, suggesting that heave ways will be a problem there. \textbf{C} and \textbf{D} show the consequences of these temperatures according to our model, with strong effects at higher temperatures but Mexico City, the capital and largest city, is spared of the worst effects because of its temperate climate.}
\label{fig6}
\end{figure*}

We explore the same projections in Mexico, which show the same strong effects from extreme heat and mild effects from temperatures that will occur often. An important difference to note is that while the Southern, tropical areas in Mexico will have the highest average temperature, with Fig.\textbf{\ref{fig6}}A showing most days over $35^{\circ}$UTCI by 2050, but the desert near the Northern border—where much of the country's manufacturing happens—will experience more heat waves, highlighted in Fig. \textbf{\ref{fig6}}B. Because the literature indicates some adaptation to heat, these heat waves in the North could disrupt activity more than stably higher temperatures in the South; although heat waves will be less common in temperate regions, these areas are most at risk of disruption when heat waves do occur. 

In Fig. \textbf{S8}, we show similar results for Indonesia, with many days over $35^{\circ}$UTCI in Sumatra and Borneo. While temperatures of $40^{\circ}$UTCI will be rarer, shown in \textbf{B} they are forecast to happen despite never appearing in our 2019 data. Because Indonesia is in the tropics and does not experience much temperature variability, our model projects strong effects for heat that exists beyond the current distribution. Indonesia is planning to move its capital to the island of Borneo, from Java, by 2045. While this megaproject intends to curb the effects of subsidence, temperatures in this new region, according to CMIP6 data, will experience more hot days than in the existing capital region. Shown in Fig. \textbf{S8C} and \textbf{D}, the consequences of these temperatures, according to our modeling, are similar to the effects we see in India: small cities areas with larger relative effects and larger urban areas with larger aggregate disruptions. 

A key caveat in year-round modeling: warming formerly cold periods can boost activity, so in India substituting 2050 weather yields only a 1\% drop overall. Because Indonesia does not have cold weather, rising temperatures there are associated with an decline of 2\% across the year. In Mexico, activity will fall as temperatures rise, by 3\% on aggregate. This points to changing dynamics: we can expect heat waves to increase in both frequency and magnitude, and thus for associated disruptions to increase, but rising temperatures will have mixed effects. These effects will be spatially and temporally heterogeneous, affecting cities and seasons in different ways.  

\section*{Discussion}
Heat waves are expected to become a regular occurrence in urban areas over the coming decades. While efforts to mitigate climate change are important, we must also develop adaptive strategies. Understanding the relationship between human behavior and extreme weather, such as heat waves, is essential for this. Our study considerably expands to a growing body of research by examining how extreme heat affects human mobility in three large developing countries. We findings show that heat waves significantly change urban mobility patterns. By examining three diverse counties across different climate zones, our research allows for comparisons and helps identify broader behavioral trends. The highest activity levels and longest trips occur at the mean temperatures for each region. However, the specifics vary between these counties. For instance, in India, activity levels were more affected by extreme high temperatures than by lower ones, whereas in Mexico and Indonesia, the reduction in activities was almost symmetric between high and low temperature extremes. 

Our results align with existing research on the economic impacts of high temperatures \cite{somanathan2021impact, schlenker2009nonlinear, zhang2018temperature} which found that productivity losses are the result of lower crop yields or impaired worker performance; we expand on this by showing that extreme heat also influences travel decisions. For example, anecdotal evidence suggests that farm workers avoid outdoor labor when temperatures exceed $40^{\circ}$UTCI and even service workers need to change transportation modes to reduce heat stress from walking \cite{cornish2024learning}, and our data confirms this while pointing to broader changes cities. Research on congestion indicates that people are willing and able to shift trips away from congested periods in the presence of taxation \cite{kreindler2024peak}; in showing that people substitute afternoon activity for morning activity, we find that heat functions as a tax on activity that people can avoid.   

In India and Mexico, respectively, we observe a noticeable decline in activity levels when average temperatures exceed 33$^{\circ}$UTCI and 35$^{\circ}$UTCI. We use rolling averages to show that, in India, while longer heat waves have stronger effects, activity resumes when high temperatures persist for a week or more. While extreme heat never shows a positive effect on mobility in our data, the negative effect converges to 0 when the temperature averages more than $40^{\circ}$UTCI for 6 or 7 days. We are unable to examine the mechanisms by which this occurs but if it is the result of individuals needing to return to work, it adds to the importance of cooling efforts. 

However, the effects we observe are not equally distributed. They are more pronounced in economically disadvantaged areas, indicating that these regions are more vulnerable to extreme weather disruptions. We find in India that foot traffic in areas characterized by high deprivation fell faster at higher temperatures than foot traffic in those with low deprivation. Changes in activity will result from combinations of forces: the ability of the wealthy to keep cool with air conditioning, for instance, might buoy activity in wealthy areas while pressures to earn might sustain activity in poor areas. Our findings suggest that for the poor, the expected gains from working are less than the dangers from working at high temperatures. This is especially concerning because residents in these areas are more likely to hold jobs that require physical labor, making adaptation more difficult. 

A limitation in our study is the coverage of the mobile phone data that we use. The data are sparser in India than they are in Mexico, with Indonesia in between, and compensate for potential biases by filtering our data to cities, where coverage is best. In Mexico, where we are most confident in the data, we see the strongest effects and the cleanest differentiation across the temperature gradient in both the GAM and TWFE models, lending credibility to our estimates. That the data in each country suggest an optimal temperature is also encouraging, but the future work will need to triangulate these findings. 

Finally, by modeling future climate scenarios, we estimate how human activity will be affected in the years to come. These projections allow us to better understand the evolving situation, not just due to changing weather patterns but also due to broader shifts, such as the planned relocation of Indonesia’s capital from Java to the island of Borneo. Heat waves in India's Rajasthan and Indonesia's Sumatra will have the strongest effects on mobility, with temperatures exceeding $40^{\circ}$UTCI more than 100 days per year in both, changes that are associated with 10\% declines in activity according to our models.  

An remaining question is whether or not populations will gradually adapt with rising temperatures. Some places will have stronger heat waves, like Mexico's arid North, and others will have higher but stabler temperatures, like its tropical South. Recent work on economic growth and public health looking across 5 decades suggests the answer is no: for most variables, responses to extreme heat have not changed \cite{burke2024we}. An important finding in our work is that more deprived areas see stronger effects than less deprived ones, which has implications for adaptation: these areas will be amongst the last to adopt air conditioning. In the coming decades, our results may hold for vulnerable populations even if others are able to adapt. 

\section*{Methods}
We use origin-destination matrices for three nations—India, Indonesia and Mexico—spanning 2019. These data contain flows between tessellated units, called geohashes. The geohashing system creates a nested hierarchy wherein longer codes indicate sharper resolution. To preserve the anonymity of users, all data have been aggregated by the data provider to the geohash5 level—with cells that are on average $\sim22\text{km}^2$—and do not include any individual records. Flows are summed from trips between any geohash6 cell where a device has recurring GPS signals \cite{zhang2024netmob2024}, and can thus occur in the data within the same origin and destination geohash5 as a loop. Aggregated data were provided by Spectus Social Impact as part of the Netmob 2024 conference. Data are collected with the informed consent of anonymous users who have opted in to anonymized data collection for research purposes. 

For each geohash5 in the data, using Google Earth Engine \cite{gorelick2017google} we also gather 365 days of weather data from the ERA5 daily climate aggregates \cite{ERA5}. These data gives us temperature, in the form of Universal Thermal Climate Index (UTCI), per day per geohash5, and it allows us to observe differences within and between regions without relying on weather station data. We acquire modeled estimates of these variables for each day of 2050 using spatially disaggregated data from the Coupled Model Intercomparison Project \cite{thrasher2022nasa}, an international collaboration to produce estimates of Earth's climate through 2100. These datasets both resolve to $\sim775\text{km}^2$, and thus also capture spatial heterogeneity in weather conditions. Finally, we attach data—via zonal statistics—on population \cite{schiavina2023ghs} and deprivation \cite{ciesin} to each geohash5, using gridded measures estimated from remote sensing. 

Using these data, we perform further checks on these data to understand is coverage and reliability in \textbf{S2}; while the number of devices in a given cell shows a strong correlation with the number of residents in it, we also see that more deprived areas are less represented in the data. Most of this bias comes from cells with very low population, which are also typically poor, so we clip our data to ``functional urban areas'' \cite{schiavina2023fua}. (We use multiple filters, show in  \textbf{\ref{T2}}, to stress test our models, but in the main analysis we use the ``low'' constraint filter.) 

\subsection*{Modeling the causal effect}
We employ a two-way fixed effects (TWFE) approach to model the relationship between temperature anomalies and mobility in a given geohash5 cell, specified as

\begin{equation}
\ln(y_{it})=\beta_0+\beta_1 \Theta_{it}+\varphi_i+\nu_t+\varepsilon_{it}
\end{equation}

where $y_{it}$ represents mobility (measured as the number of trips ending in cell $i$ on day $t$). The independent variable $\Theta_{it}$ captures temperature stress in cell $i$ on day $t$, which we define as the deviation from the cell's mean temperature over the study period. This specification allows us to examine how departures from typical local conditions affect mobility patterns, rather than absolute temperature levels which could conflate adaptation effects.
The model includes two types of fixed effects: $\varphi_i$ represents cell-specific fixed effects that control for time-invariant characteristics of each geographic unit, such as infrastructure, elevation, or proximity to water bodies. $\nu_t$ captures day fixed effects that account for temporal factors affecting all cells simultaneously, including holidays, weekends, and seasonal patterns. These day fixed effects also help control for potential sampling variations in the GPS data collection, as the number of active devices may fluctuate based on the set of applications providing data on any given day.

To account for potential heterogeneity in temperature effects across different times of day, we estimate separate models for morning (6:00-12:00), afternoon (12:00-18:00), and evening (18:00-24:00) periods. This temporal disaggregation reveals how mobility responses to heat stress vary throughout the day, capturing potential behavioral adaptations such as shifting activities to cooler hours.

We cluster standard errors at the cell level to account for potential serial correlation in mobility patterns within geographic units. The coefficient of interest, $\beta_1$, represents the semi-elasticity of mobility with respect to temperature stress, interpretable as the percentage change in mobility associated with a one-unit increase in temperature deviation from the local mean. 

\subsection*{Modeling the temperature gradient}
We begin using a generalized additive model (GAM) to measure the partial effect of temperature at different levels \cite{wood2001mgcv}. We build a directed acyclic graph (DAG) to determine what controls are necessary in our model, shown in Fig. \textbf{S3A}. In addition to temperature, solar radiation and precipitation could also modulate activity, with rain dampening activity and sun heightening, so we control for these along with day-of-year. Day-of-year is modeled with a cyclic cubic spline, which captures seasonality by allowing for variation by day while forcing the value of the spline at the start to equal its value at the end. Because geography and area characteristics will also influence activity, we use various geohash fixed effects, allowing the intercept to vary according to the unique activity profile of each geographic unit. A key assumption with this approach is that the response curve is the same across space—only the intercept will vary. Together, these spatial and temporal controls constitute the minimally sufficient adjustment set. While they are not confounds, adjusting for day-of-week and holiday effects improves the precision of our estimates. The resulting equation is  

\begin{equation}
T = \beta_0 + f_1(x_1) + f_2(x_2) + \mathbf{X}\boldsymbol{\beta} + \varepsilon
\label{eq1}
\end{equation}

\noindent
where $T$ is the number of trips ending in the geohash5, $f_1(x_1)$ represents the smooth function of temperature, $f_2(x_2)$ represents the smooth function of day (with a cyclic cubic spline), $\mathbf{X}$ is a matrix of control variables, and $\boldsymbol{\beta}$ is a vector of coefficients for the control variables. $\varepsilon$ is the error term. 

\subsection*{Modeling the time series}
Because GAMs do not have an explicit time component, we then model the time series as a robustness check. To analyze the relationship between activity and temperature while accounting for temporal dependencies, we employ an Autoregressive Integrated Moving Average model with Exogenous variables (ARIMAX) \cite{hyndman2008automatic}. This model combines autoregressive terms, moving average components, and exogenous factors to capture the complex dynamics of the time series. Specifically, our model incorporates AR(1) and AR(7) terms to account for immediate changes and day-of-the-week patterns in activity. The moving average component allows the model to consider the impact of past shocks and manages seasonality. Temperature is included as an exogenous factor with both linear and quadratic terms to capture potential nonlinear effects on activity. Our model is specified as

\begin{equation}
T_t = c + \phi_1 T_{t-1} + \phi_7 T_{t-7} + \beta_1 \text{Temp}_t + \beta_2 \text{Temp}_t^2 + \theta_1 \varepsilon_{t-1} + \varepsilon_t
\label{eq2}
\end{equation}

\noindent
where $T_t$ represents economic activity measured as Trips at time $t$, $UTCI$ is a constant term, $\phi_1$ and $\phi_7$ are autoregressive coefficients, $\beta_1$ and $\beta_2$ are coefficients for linear and quadratic temperature effects, $\text{Temp}_t$ is the temperature at time $t$, $\theta_1$ is the moving average coefficient, and $\varepsilon_t$ is the error term. This model structure allows us to simultaneously account for time-dependent patterns in economic activity and the influence of temperature, providing a comprehensive framework for analyzing the complex relationships within our data.

We compute the average coefficient for $\text{Temperature}^2$, which indicates the degree to which the effect of temperature curves at extremes, and we identify the ideal temperature by making predictions using the model for each time series at all temperatures and observing the point at which predictions begin to fall. We run a sensitivity analysis because not all time series span the full length of the year, showing the average coefficient at various minimum thresholds.    

\subsection*{Projecting into the future}
In order to understand how rising temperatures will impact behaviors, we take two approaches. First, we take the covariates from Eq. \ref{eq1} but switch the temperatures on any given day for temperatures from the same day-of-year in 2050 using CMIP6 SSP 8.5 \cite{thrasher2022nasa}. This gives us an aggregate estimate, incorporating both warmer winters (which might be good) and hotter summers (which might be bad). Second, we explore specific temperature thresholds that our modeling has shown to exert strong effects on mobility and explore how often these thresholds will be crossed according to CMIP6 SSP 8.5 estimates in 2050. To see estimates of temperatures in our 3 study countries through 2100, see Fig. \textbf{S9}.


\bibliography{heat}

\section*{Acknowledgements}
The authors would like to thank the Spectus and the NetMob team for making the data available, as well as Wenlan Zhang for responding to questions and updating the data on multiple occasions.

\section*{Author contributions statement}
\textbf{A.R.} Conceptualization, methodology, investigation, writing, reviewing, editing; \textbf{O.H.} Conceptualization, methodology, investigation, writing, reviewing, editing; \textbf{P.B.} Conceptualization, methodology, investigation, writing, reviewing, editing. 

\subsection*{Data and code availability}
Mobility data are available on request as part of the NetMob 2024 Data Challenge \cite{zhang2024netmob2024}. ERA5 climate recordings and CMIP6 climate projections are accessible via Google Earth Engine. All code is available upon request. 
\section*{Competing interests}
The authors declare no conflict of interest.

\end{document}


\flushbottom
\maketitle

\tableofcontents

\clearpage


\section{Network properties}

\begin{figure*}[h!]
\centering
\includegraphics[width=1\textwidth]{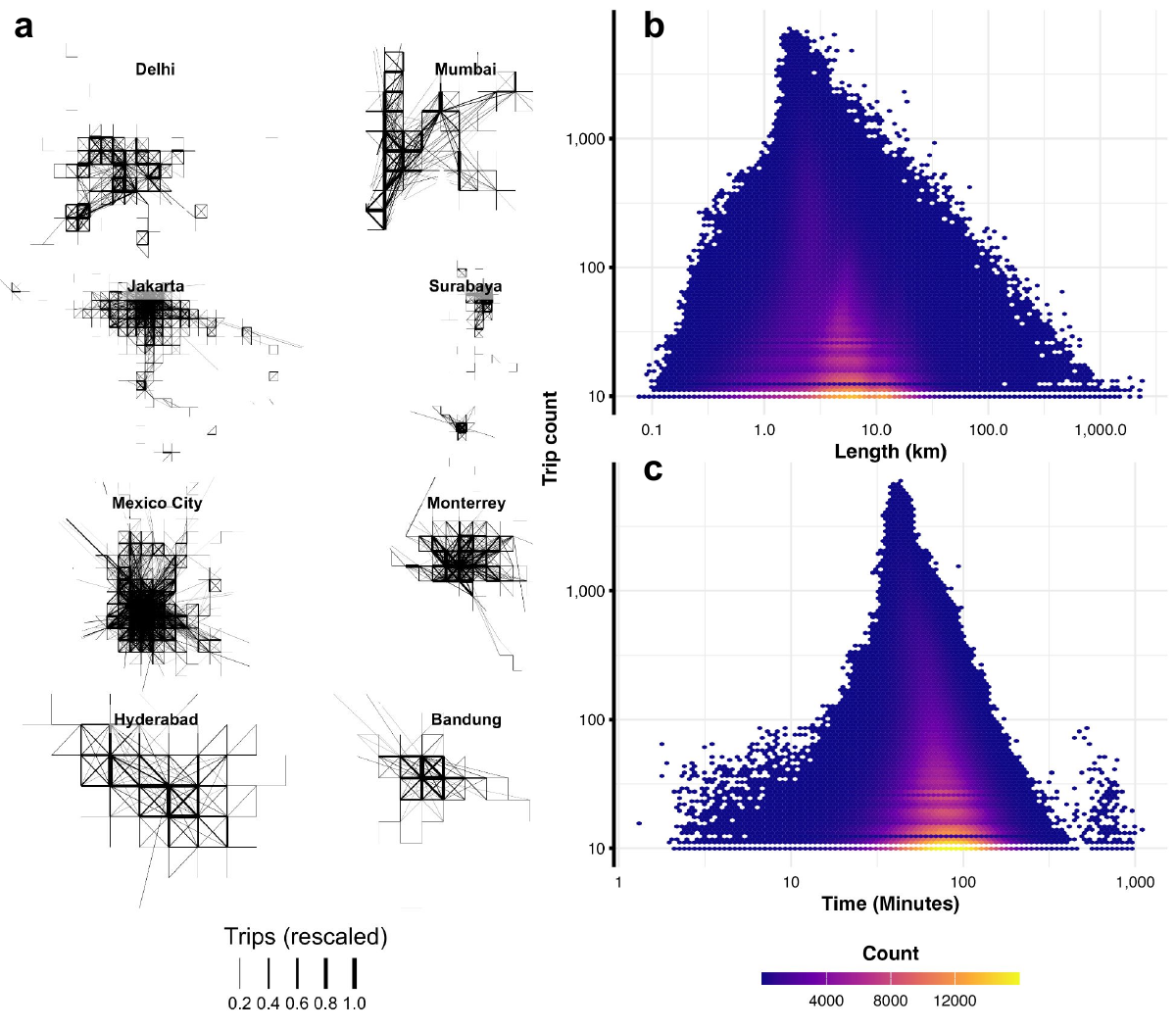}
\caption{\textbf{Networks and statistics.} \textbf{A} We construct networks where edges correspond to trips and nodes corresponds to geohash5 aggregations. \textbf{B} and \textbf{C} show the properties of these edges, with trip times averaging between 50 and 100 minutes and lengths in kilometers averaging between 5 and 10. We note a distance and time decay, where trips of long lengths and times are less frequent across the sample.}
\label{S1}
\end{figure*}

\clearpage

\begin{figure*}[h!]
\centering
\includegraphics[width=1\textwidth]{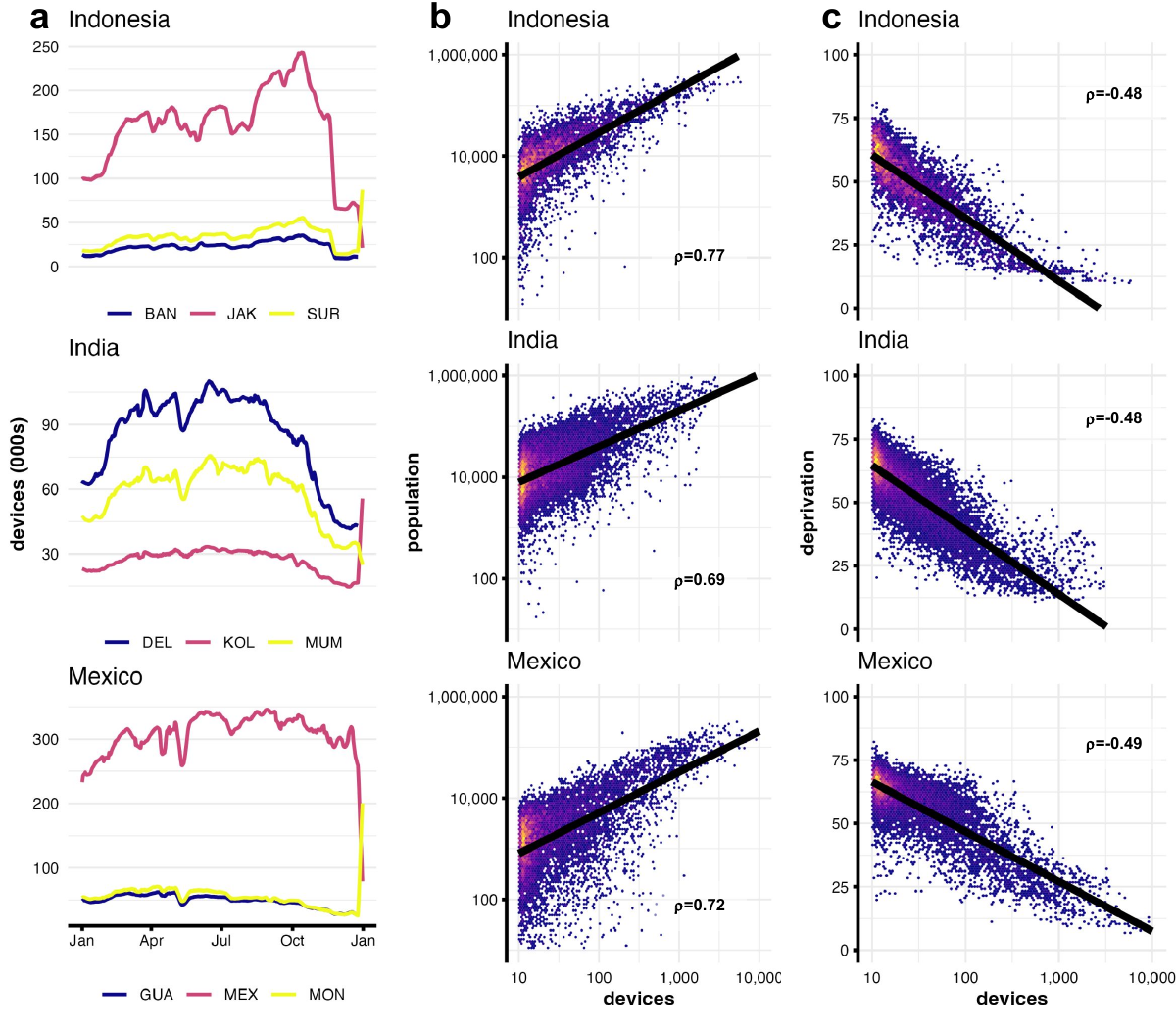}
\caption{\textbf{Baises in the data.} \textbf{A} We show the time series for various top cities in the data, finding irregularities in the beginning and end of the year, which we remove from the data. We also follow the recommendation the data provider \cite{zhang2024netmob2024} and remove a number of other anomalous days in May and August. \textbf{B} and \textbf{C} show how device counts in each geohash correlate with population and deprivation. While see a strong correlation to population by geohash in our data, lending confidence to our estimates, we also see that more deprived areas are less represented in the data. Because population itself is anticorrelated with deprivation ($\rho=-0.7$), this is not entirely the product of bias, but we do our best to address this with stratification and sensitivity analysis.}
\label{S2}
\end{figure*}

\clearpage

\begin{table}[h!]
\begin{center}
\begin{tabular}{l|r|r}
\textbf{Origin-Destination Patterns} & \textbf{Trips} & \textbf{Share (\%)} \\ \midrule
\multicolumn{3}{l}{\textit{India}} \\ \midrule
Urban-rural & 4,527 & 0.01\% \\
Rural-urban & 4,870 & 0.01\% \\
Inter-urban & 30,152 & 0.05\% \\
Rural-rural & 2,333,024 & 3.54\% \\
Within-urban & 63,527,103 & 96.40\% \\
\hspace{0.3cm} \textit{of which within-cell} & 52,383,732 & 79.49\% \\
Total trips & 65,899,676 & 100.00\% \\ \midrule
\multicolumn{3}{l}{\textit{Indonesia}} \\ \midrule
Urban-rural & 6,952 & 0.02\% \\
Rural-urban & 8,234 & 0.02\% \\
Inter-urban & 16,182 & 0.04\% \\
Rural-rural & 1,265,869 & 3.04\% \\
Within-urban & 40,366,663 & 96.89\% \\
\hspace{0.3cm} \textit{of which within-cell} & 30,211,502 & 72.51\% \\
Total trips & 41,663,900 & 100.00\% \\ \midrule
\multicolumn{3}{l}{\textit{Mexico}} \\ \midrule
Urban-rural & 221,369 & 0.17\% \\
Rural-urban & 226,830 & 0.18\% \\
Inter-urban & 129,046 & 0.10\% \\
Rural-rural & 15,516,385 & 12.06\% \\
Within-urban & 112,561,977 & 87.49\% \\
\hspace{0.3cm} \textit{of which within-cell} & 76,076,308 & 59.13\% \\
Total trips & 128,655,607 & 100.00\%
\end{tabular}
\caption{\textbf{Comparison of trip patterns across all three countries.} Most trips in all countries are urban-urban and very few begin or end in rural areas. In Mexico, the most developed country in our dataset, the share of trips originating and terminating the same cell is the lowest, indicating greater mobility.}
\label{T1}
\end{center}
\end{table}

\clearpage

\section{Alternative specifications}

\begin{figure*}[h!]
\centering
\includegraphics[width=1\textwidth]{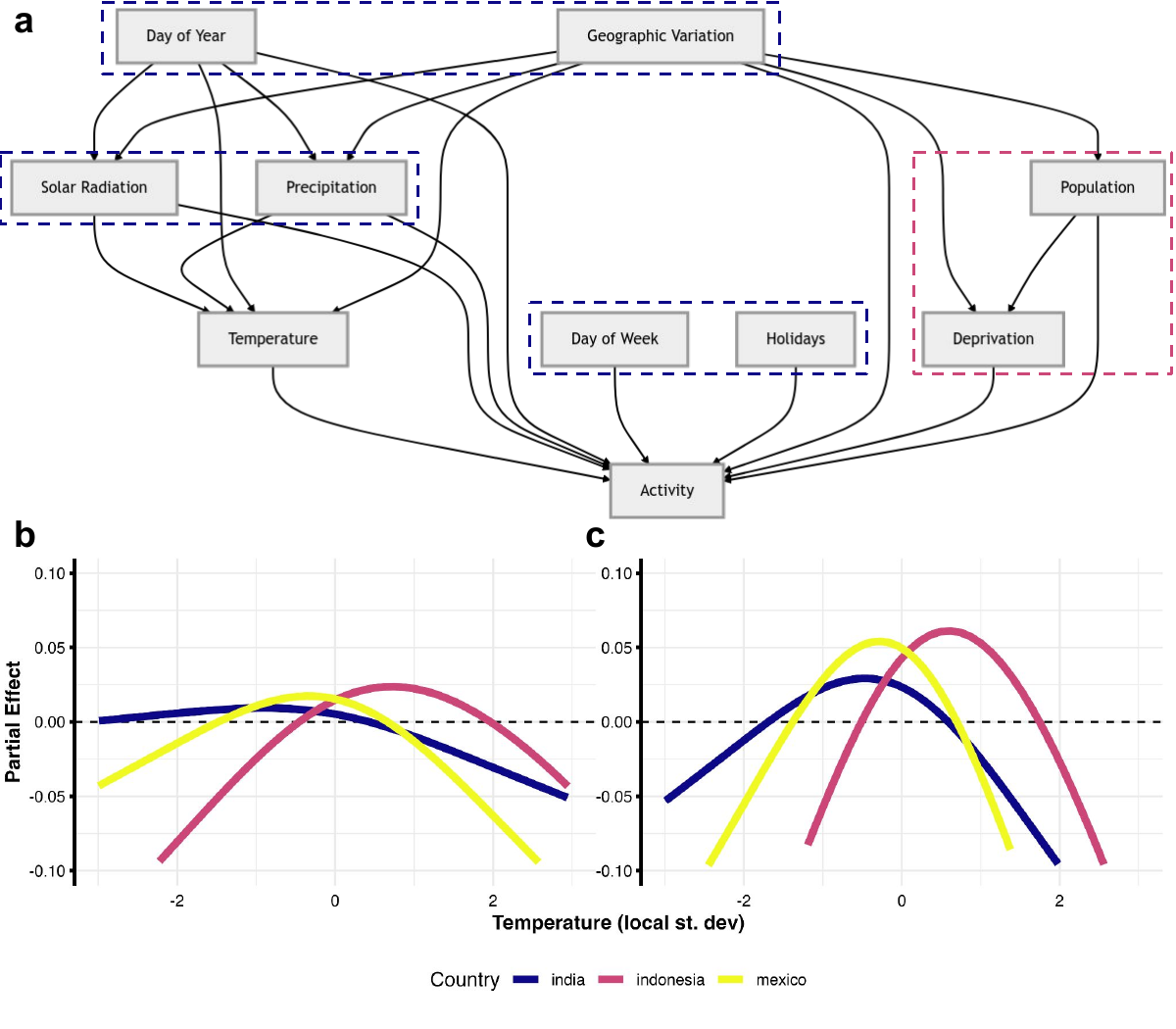}
\caption{\textbf{Modeling temperature and activity.} \textbf{A} We set up a directed acyclic graph (DAG) to ensure that we close all necessary causal paths to activity; we control for day-of-year, day-of-week, holidays, solar radiation, precipitation, and use geographic fixed effects that necessarily stratify by population and deprivation. \textbf{B} Model results for different countries, showing that the highest activity levels occur at average temperatures for an area, and that high extremes correspond with fewer trips. \textbf{C} Model results using trip duration rather than trip count reveal that the longest trips tend to occur at average temperatures, with extreme high temperatures leading to shorter trips.}
\label{S3}
\end{figure*}

\clearpage

\section{Placebo tests}

\begin{figure*}[h!]
\centering
\includegraphics[width=1\textwidth]{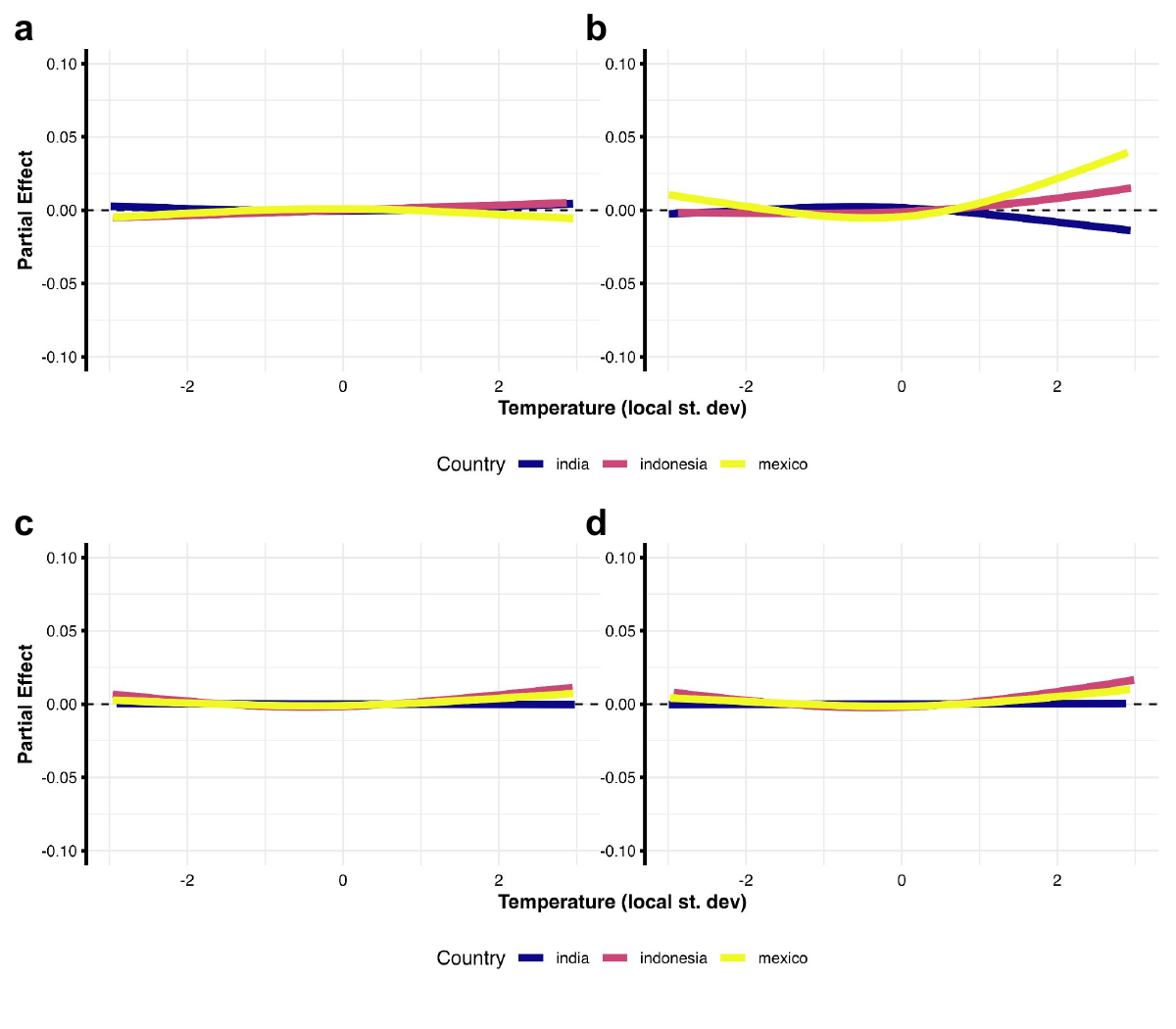}
\caption{\textbf{Permutations tests.} We use a variety of permutation tests to ensure that the relationship we observed is not spurious, including shuffled temperatures within GeoHash5 between dates, giving each cell a temperature from a different time of year in that cell. \textbf{A} and \textbf{B} does this for duration and trip count, respectively. We also shuffled between GeoHash5 within dates, giving each cell a new temperature from the same date in a different cell, again for duration and trip count in \textbf{C} and \textbf{D}.}
\label{S4}
\end{figure*}

\clearpage

\section{Sensitivity analysis}

\begin{table}[h!]
\begin{center}
\begin{tabular}{l|r|r}
\textbf{Constraint Level and Requirements} & \textbf{No. of cells} & \textbf{No. of cities} \\ \midrule
\multicolumn{3}{l}{\textit{India}} \\ \midrule
\textit{Low constraint:} & 1,259 & 482 \\
Min. days with mobility=50 & & \\
Min. average mobility=10 & & \\ \midrule
\textit{Medium constraint:} & 1,043 & 384 \\
Min. days with mobility=100 & & \\
Min. average mobility=15 & & \\ \midrule
\textit{High constraint:} & 832 & 262 \\
Min. days with mobility=150 & & \\
Min. average mobility=20 & & \\ \midrule
\multicolumn{3}{l}{\textit{Indonesia}} \\ \midrule
\textit{Low constraint:} & 701 & 199 \\
Min. days with mobility=50 & & \\
Min. average mobility=10 & & \\ \midrule
\textit{Medium constraint:} & 574 & 161 \\
Min. days with mobility=100 & & \\
Min. average mobility=15 & & \\ \midrule
\textit{High constraint:} & 461 & 103 \\
Min. days with mobility=150 & & \\
Min. average mobility=20 & & \\ \midrule
\multicolumn{3}{l}{\textit{Mexico}} \\ \midrule
\textit{Low constraint:} & 964 & 177 \\
Min. days with mobility=50 & & \\
Min. average mobility=10 & & \\ \midrule
\textit{Medium constraint:} & 890 & 176 \\
Min. days with mobility=100 & & \\
Min. average mobility=15 & & \\ \midrule
\textit{High constraint:} & 731 & 109 \\
Min. days with mobility=150 & & \\
Min. average mobility=20 & & \\
\end{tabular}
\caption{\textbf{Sample sizes under varying data quality constraints.} We test our results on various samples of the data, conditioning on the level of mobility in the cells to ensure that large changes on small values are not driving our results.}
\label{T2}
\end{center}
\end{table}

\clearpage

\begin{figure*}[h!]
\centering
\includegraphics[width=1\textwidth]{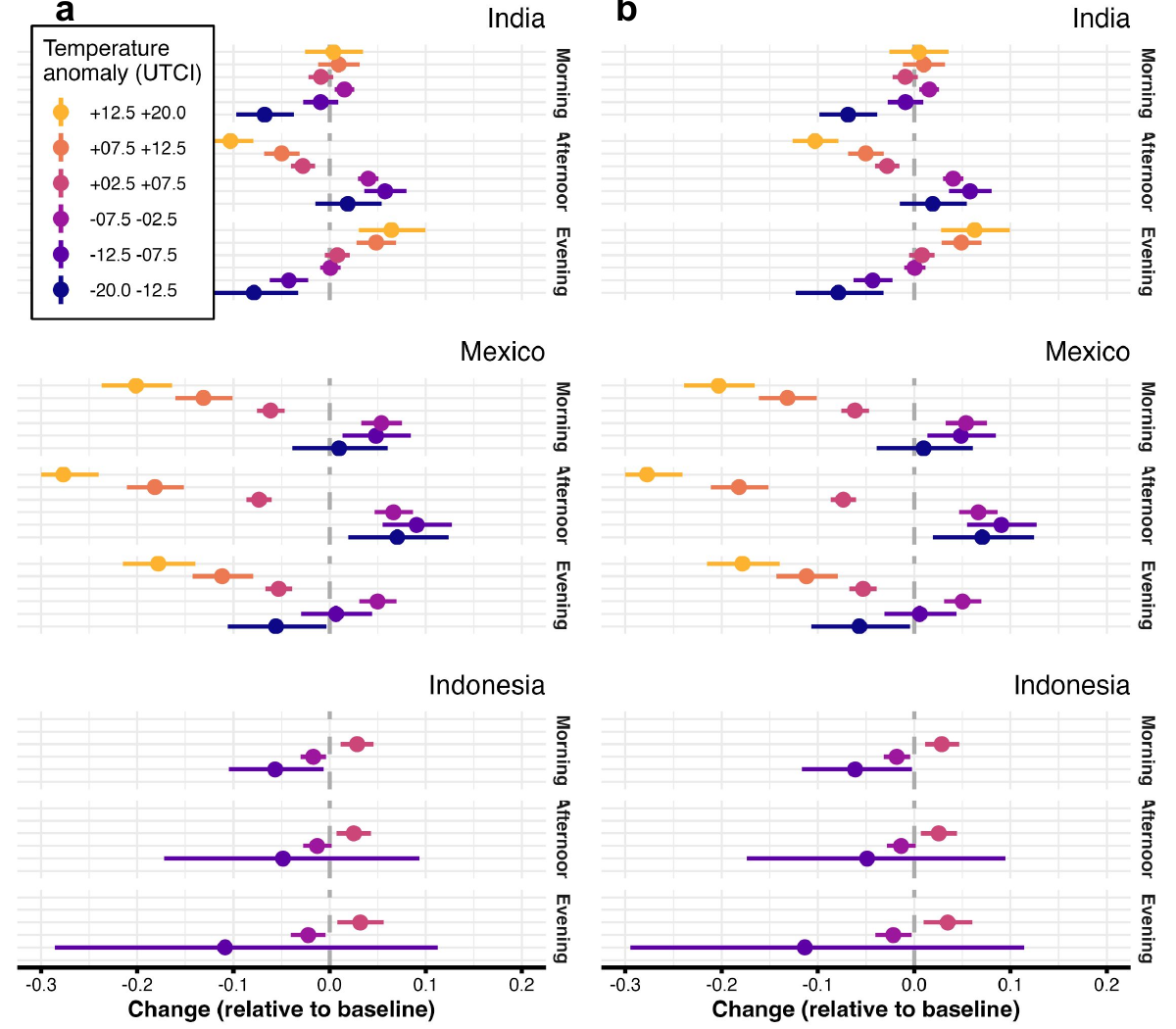}
\caption{\textbf{Medium and high constraints.} We show the results of our TWFE regressions for both \textbf{A} medium and \textbf{B} high constraints respectively. These results, which are almost identical, show that different subsets of the data behave in similar ways, and our results are not driven by certain groups of cells with high rates of change on low values.}
\label{S5}
\end{figure*}

\clearpage

\section{Time-of-day and population}

\begin{figure*}[h!]
\centering
\includegraphics[width=1\textwidth]{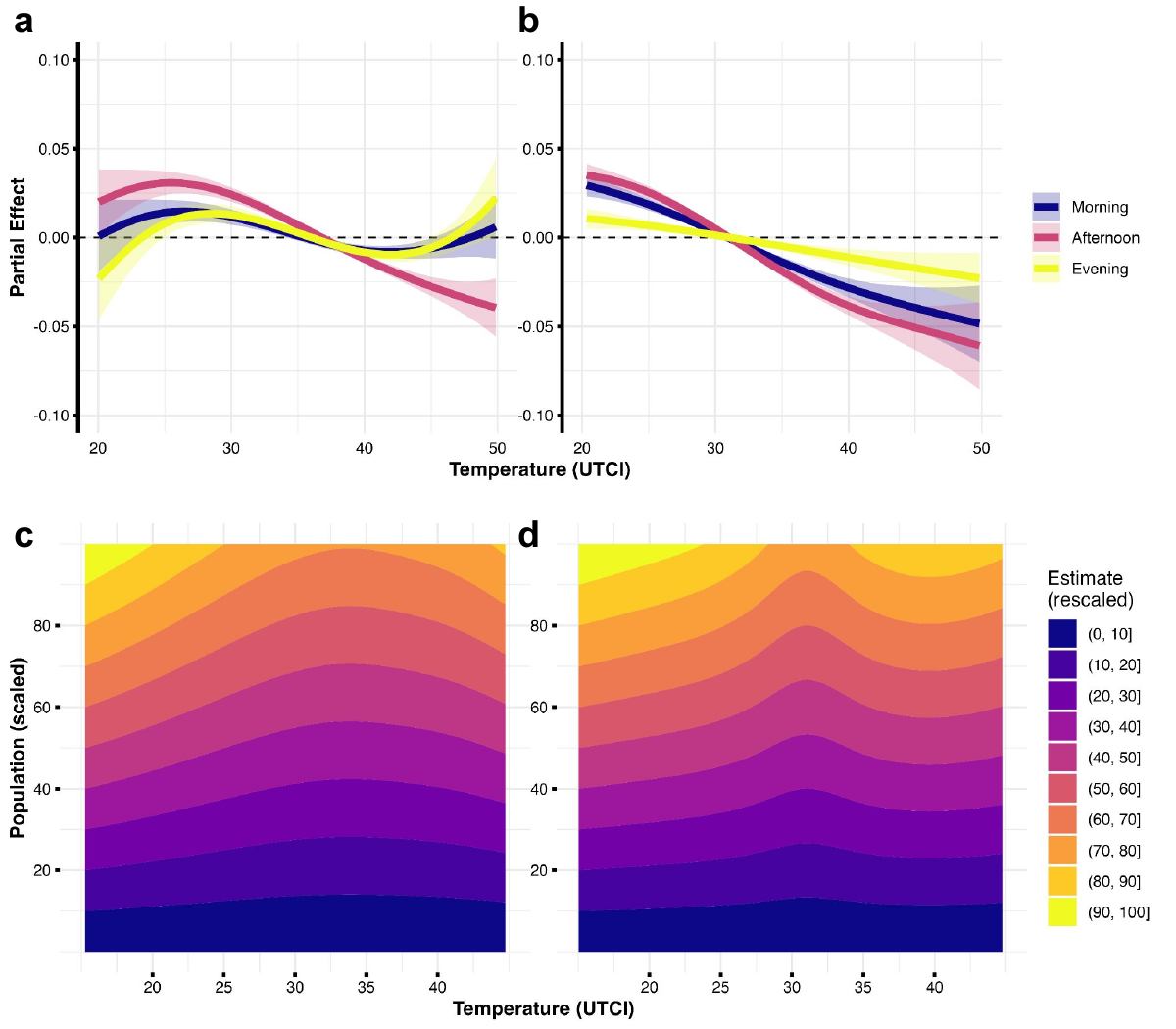}
\caption{\textbf{Mobility by time of day and population.} \textbf{A} When we decompose the effect of temperature on activity by time of day, we see that the reduction in activity comes during the afternoons, suggesting that people are avoiding the hottest part of the day.  \textbf{B} In Mexico, all periods of the day see reductions in activity. We see that in India \textbf{C} and Mexico \textbf{D}, larger populations see stronger temperature effects in absolute terms, with higher predicted values and moderate temperatures and lower predicted values at the extremes, while these curves are flattened for smaller populations.}
\label{S6}
\end{figure*}

\clearpage

\section{ARIMAX curves}

\begin{figure*}[h!]
\centering
\includegraphics[width=1\textwidth]{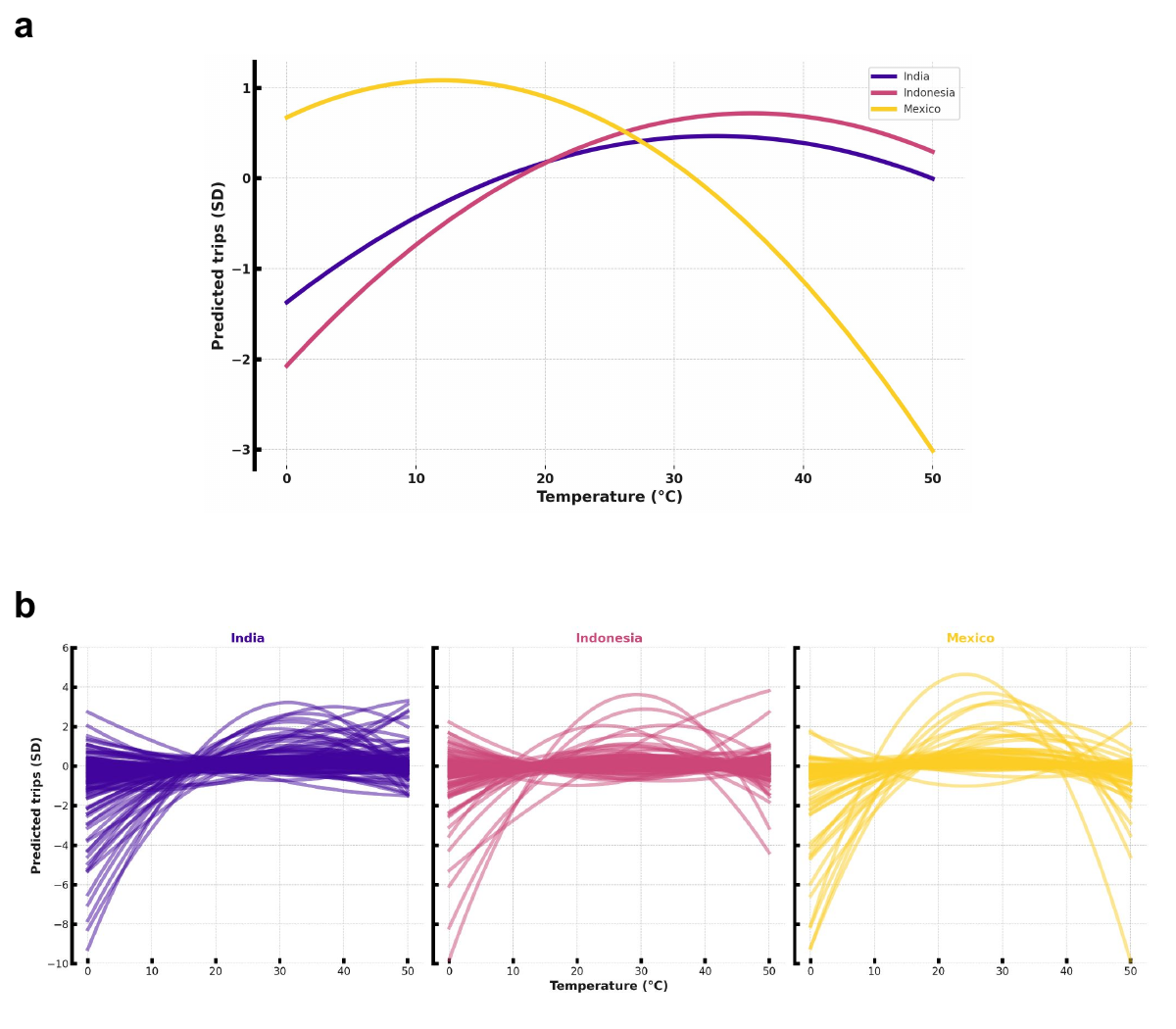}
\caption{\textbf{ARIMAX curves.} \textbf{A} We plot the curves for optimal temperature given the coefficients from the average fitted ARIMAX model, seeing that the predictions turn negative between $30$ and $40^{\circ}$C for both India and Indonesia, but fall much earlier in Mexico—which has a temperate climate many populous areas like Mexico City and Guadalajara. \textbf{B} We disaggregate those averages to show all curves from all geohash3 ARIMAX models: there are some curves that show a positive effect of temperature but most show a negative one, in particular those fit in India.}
\label{S7}
\end{figure*}

\clearpage

\section{Projections for Indonesia}

\begin{figure*}[h!]
\centering
\includegraphics[width=1\textwidth]{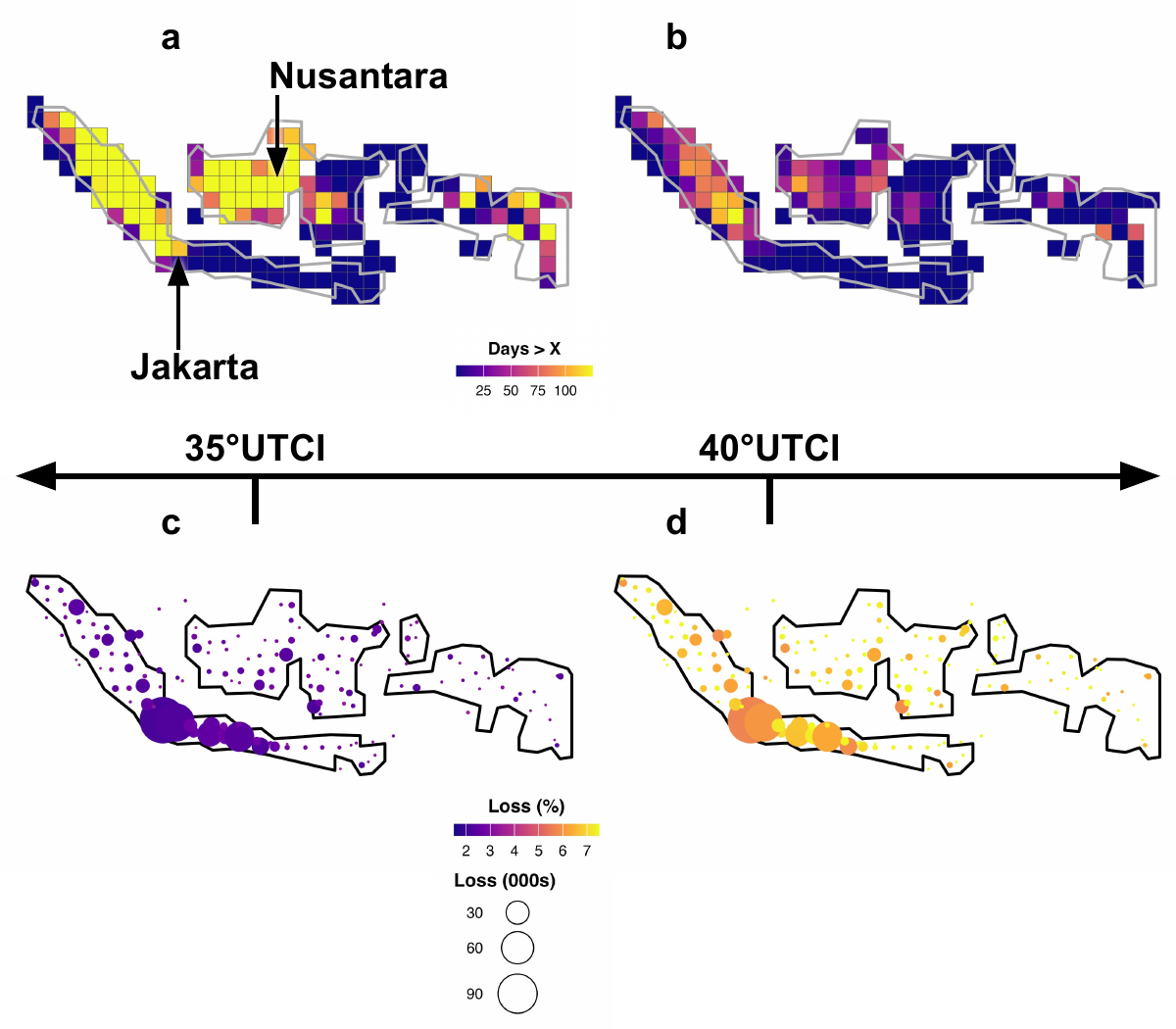}
\caption{\textbf{Temperatures and activity in 2050.} \textbf{A} and \textbf{B} show frequency of days above $35^{\circ}$UTCI and $40^{\circ}$UTCI, respectively; we see that Borneo and Sumatra will have more frequent days above $35^{\circ}$UTCI while days $40^{\circ}$UTCI will be concentrated in Sumatra, suggesting that heave ways will be a problem there. We note here that Indonesia is in the process of moving its capital city to a hotter part of the country, which will either force adaptation for new residents or limit activity. \textbf{C} and \textbf{D} show the consequences of these temperatures according to our model, with strong effects at higher temperatures but Indonesia, the existing capital and largest city, is will show the largest effects in absolute terms.}
\label{S8}
\end{figure*}

\clearpage

\section{CMIP6 estimates under different scenarios}

\begin{figure*}[h!]
\centering
\includegraphics[width=1\textwidth]{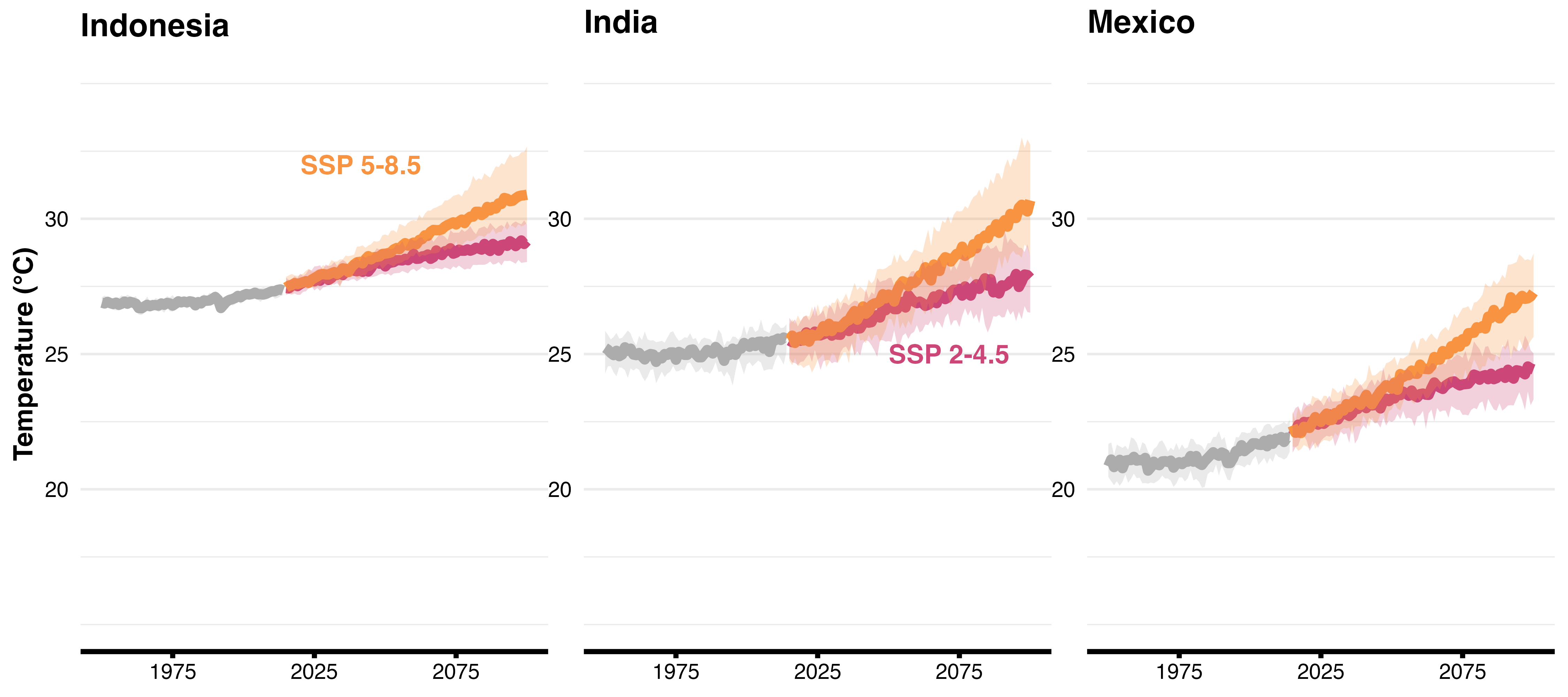}
\caption{\textbf{Temperatures under SSP 4.5 and 8.5 through 2100.} We show CMIP6 estimates under different scenarios for each country, with Mexico seeing warming from a low average today, India seeing the strongest change, and Indonesia seeing less of a change but from a high average today. These estimates are the averages from ensembles of models and the ribbons indicate the 10th and 90th percentiles of all models \cite{WorldBank_CCKP}. }
\label{S9}
\end{figure*}

\clearpage

\clearpage

\bibliographystyle{naturemag}
\bibliography{heat}